\documentclass[11pt]{article}
\usepackage{moriond,epsfig,color}

\def\be{\begin{equation}}
\def\ee{\end{equation}}
\def\bea{\begin{eqnarray}}
\def\eea{\end{eqnarray}}

\begin{document}
\hfill PITHA-05-08
\vspace*{4cm}
\title{CKM Overview and Determinations from $\boldmath B\unboldmath$ Decays}

\author{Stefan W. Bosch}

\address{Institut f\"ur Theoretische Physik E; RWTH Aachen\\ D-52056 Aachen, Germany}

\maketitle\abstracts{
This talk gives an introduction to the Standard Model Cabibbo-Kobayashi-Maskawa matrix and methods to extract information on the parameters of the unitarity triangle from $B$ decays.}

\section{Introduction}\label{intro}
Flavour physics is at the moment the most active field of elementary particle physics. In particular the dedicated $B$-physics experiments BaBar and Belle produce a wealth of data which allows for a determination of fundamental parameters of the Standard Model with unprecedented accuracy. Flavour physics and CP violation are governed by the Cabibbo-Kobayashi-Maskawa (CKM) matrix\cite{CKM} which relates the flavour and mass eigenstates of quarks. The CKM-matrix elements appear as coupling constants in the charged-current transitions.

As a unitary complex $3\times 3$ matrix it has in principle nine real parameters, five of which can be eliminated due to phase redefinitions of the quarks. The three-generation CKM matrix therefore has three angles and one complex phase. The latter one is the only source of CP violation within the Standard Model (SM). There are many different ways of parametrizing the CKM matrix. For practical purposes most useful is the so called Wolfenstein parametrization\cite{Wolfen}
\begin{equation}
\label{wolfen}
V_\mathrm{CKM}=\left(
  \begin{array}{ccc} 
    1-\frac{\lambda^2}{2} & \lambda & A\lambda^3 (\rho-i\eta) \\
    -\lambda & 1-\frac{\lambda^2}{2} & A\lambda^2 \\
    A \lambda^3 (1-\rho-i\eta) & -A\lambda^2 & 1
  \end{array}\right)
\end{equation}
which is an expansion to ${\cal O}(\lambda^3)$ in the small parameter $\lambda=|V_{us}|\approx 0.22$. It is possible to improve the Wolfenstein parametrization to include higher orders of $\lambda$.\cite{impwolf} The Wolfenstein parametrization makes the nearly diagonal structure of the CKM matrix obvious.

\begin{figure}
  \begin{center}
    \input{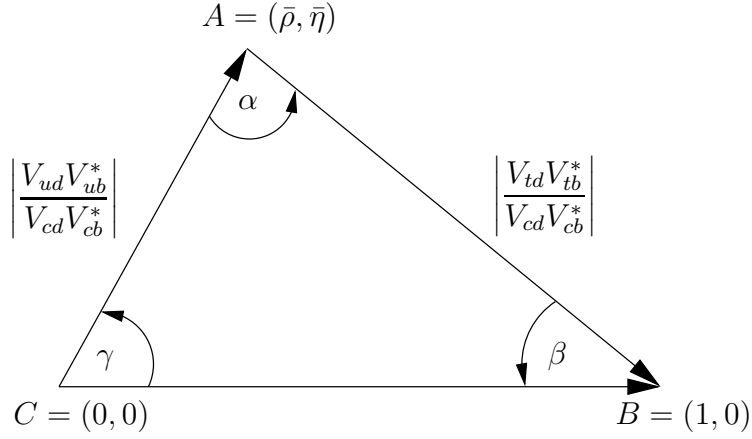}
  \end{center}
  \caption{Unitarity Triangle\label{fig:UT}}
\end{figure}

For phenomenological studies of CP-violating effects, the so called standard unitarity triangle (UT) plays a special role. It is a graphical representation of the orthogonality of the first and third column of the CKM matrix, namely
\begin{equation}
\label{ut}
V_{ud}V_{ub}^*+V_{cd}V_{cb}^*+V_{td}V_{tb}^*=0
\end{equation}
in the $(\rho,\eta)$ plane. This unitarity relation involves simultaneously the elements $V_{ub}$, $V_{cb}$, and $V_{td}$ which are under extensive discussion at present. The area of this and all other unitarity triangles equals half the absolute value of $J_{CP}=\mathrm{Im}(V_{us}^{}V_{cb}^{}V_{ub}^*V_{cs}^*)$, the Jarlskog measure of CP violation.\cite{JCP} Usually, one chooses a phase convention where $V_{cd}^{}V_{cb}^*$ is real and rescales the above equation with $|V_{cd}^{}V_{cb}^*|=A\lambda^3$. This leads to the triangle in Figure \ref{fig:UT} with a base of unit length and the apex $(\bar\rho,\bar\eta)$.
A phase transformation in (\ref{ut}) only rotates the triangle, but leaves its form unchanged. Therefore, the angles and sides of the unitarity triangle are physical observables and can be measured. The angles $\gamma$ and $\beta$ are given by the arguments of $V_{ub}^*$ and $V_{td}^*$, respectively, whereas $\alpha=\pi-\beta-\gamma$ as a consequence of unitarity. Much effort is put into the determination of all the UT parameters. One tries to measure as many parameters as possible. The consistency of the various measurements tests the consequences of unitarity in the three-generation Standard Model. Any discrepancy with the SM expectations would imply the presence of new channels or particles contributing to the decay under consideration.

All angles and sides of the standard unitarity triangle can be determined from $B$ meson decays. The most prominent determination methods are presented in the following.

\section{Unitarity triangle angles from $B$ decays}
\subsection{$\sin 2\beta$ from $B\to J/\psi K_s$}\label{beta}
When a $B$ meson decays into a final state $f_{CP}$ which is an eigenstate of $CP$, the $B$ meson can decay either directly into $f_{CP}$ or mix into a $\bar B$ before decaying. The time-dependent CP asymmetry of such a decay takes a particularly simple form:
\begin{eqnarray}
{\cal A}_{CP}(t) &=& \frac{\Gamma(\bar B^0 \to f_{CP})-\Gamma(B^0\to f_{CP})}{\Gamma(\bar B^0 \to f_{CP})+\Gamma(B^0\to f_{CP})}\nonumber\\
	&=& S_f^{\rm mix}\sin(\Delta M\, t) +C_f^{\rm decay}\cos(\Delta M\, t)
\end{eqnarray}
where $\Delta M$ is the mass difference between the mass eigenstates in the $B^0 \bar B^0$ system. The $\sin(\Delta M\, t)$ term with coefficient $S_f^{\rm mix}$ is due to interference of CP violation in mixing and decay, whereas the $\cos(\Delta M\, t)$ term with coefficient $C_f^{\rm decay}$ is non-vanishing when the decay directly violates CP. The latter is possible if at least two amplitudes with different weak and strong phases contribute to the decay under consideration. If, on the other hand, a single amplitude dominates, $C_f=0$. The prototype example of such a decay is the mode $B\to J/\psi K_s$ whose time-dependent CP asymmetry is given by $\sin(2\beta)\sin(\Delta M\, t)$ where $2\beta$ is the weak phase of $B^0 \bar B^0$ mixing.

The cleanest way to produce $B$ mesons is to operate an $e^+ e^-$ collider at the $\Upsilon (4s)$ resonance. The decay of the $\Upsilon (4s)$ produces with equal probability a $B^+ B^-$ or $B^0 \bar B^0$ pair, where the latter is produced in a coherent quantum state. To measure a time-dependent CP asymmetry in neutral $B$-meson decays, it is necessary to determine the time difference $\Delta t$ between the decays of the two mesons inside the pair. This can be achieved by measuring the distance between the two decay vertices which requires that the two $B$ mesons have a non-vanishing relative velocity. In the $\Upsilon (4s)$ rest frame the pair of $B$ mesons is produced almost at rest. In the asymmetric $B$ factories PEP-II\cite{BaBar} and KEK-B\cite{KEK-B} the electron and positron beams have unequal energies which produces the $B$ meson pairs with a boost in the laboratory frame and therefore enables the measurement of time-dependent CP asymmetries.

The current world average of $\sin 2\beta$ from $B\to J/\psi K_s$ and other charmonium modes is\cite{HFAG}
\begin{equation}\label{sinb}
	\sin 2\beta = 0.725\pm 0.037
\end{equation}
which is in perfect agreement with expectations from indirect measurements. The Standard-Model CKM picture is therefore established and the task is to look for small deviations from it.

\subsection{$\sin 2\beta_{\rm eff}$ from $b\to s$ penguins}
Decays like $B\to\phi K_s$ are also dominated by a single decay amplitude: the $b\to s$ penguin transition which in the Standard Model has no weak phase. Therefore one expects that\cite{betabs}
\begin{equation}
	|S_{\psi K_s}-S_{\phi K_s}|_{\rm SM} \leq 0.04
\end{equation}
such that the time-dependent CP asymmetry in $B\to\phi K_s$ also measures $\sin 2\beta$ to good approximation. 
The world average of $\sin 2\beta$ from $B\to \phi K_s$ and other s-penguin modes is\cite{HFAG}
\begin{equation}
	\sin 2\beta = 0.43\pm 0.07
\end{equation}
which is a $3.7\sigma$ deviation from the charmonium-mode value in (\ref{sinb}). A persisting discrepancy with more data would be a rather clean hint towards new physics. Because the charmonium modes are governed by tree level decays whereas the $b\to s$ transitions are penguin decays, new physics could easily enter $S_{\psi K_s}$ and $S_{\phi K_s}$ differently.

\subsection{$\alpha$ from $B\to \{\rho/\pi\}\{\rho/\pi\}$}
The quark-level transition $\bar b\to \bar u$ is accompanied by the CKM element $V_{ub}=|V_{ub}|e^{-i\gamma}$ which carries the weak phase $\gamma$ allowing for direct CP violation. In the interference with $B\bar B$ mixing, the resulting weak phase is $\beta +\gamma =\pi-\alpha$. The $\bar b\to \bar u$ transition appears for example in $B\to\pi\pi$, $B\to\rho\pi$, and $B\to\rho\rho$ decays. If this were the only contribution to these modes, their time-dependent CP asymmetry would directly measure the UT angle $\alpha$. However, the same final states can also be obtained via the $b\to d$ QCD-penguin diagram which carries the weak phase $\beta$. Hence, due to the interference of tree and penguin amplitudes, the time-dependent CP asymmetry in  $B\to \{\rho/\pi\}\{\rho/\pi\}$ measures some $\alpha_{\rm eff}$ whose deviation from $\alpha$ depends on the relative size of tree and penguin contributions.

Using an isospin analysis proposed first by Gronau and London\cite{GL}, it is possible to extract $\alpha$ up to discrete ambiguities constructing the triangles
\begin{eqnarray}
	A_{+-}+\sqrt{2}A_{00} &=& \sqrt{2}A_{+0}\\
	\bar A_{+-}+\sqrt{2}\bar A_{00} &=& \sqrt{2}\bar A_{+0}
\end{eqnarray}
Here the subscripts denote the charge of the final state mesons and the bar represents that a $\bar B$ instead of a $B$ meson is decaying.	 The Gronau-London method leads to a theoretically very clean determination of $\alpha$ but is experimentally extremely challenging because the small $B\to\pi^0 \pi^0$ rates and CP asymmetries have to be measured.

So far, only upper bounds on the CP-averaged branching ratios $B(B\to \pi^0\pi^0)$ and $B(B\to\rho^0\rho^0)$ exist. Grossman and Quinn showed how these measurements together with the other $B\to \{\rho/\pi\}\{\rho/\pi\}$ branching ratios can be used to get an upper bound on the error on $\sin 2\alpha$ due to penguin diagram effects.\cite{GQ} Because the bound on $B(B\to\pi^0\pi^0)$ is rather large, only the weak bound $\alpha_{\rm eff}-\alpha < 35^\circ$ can be inferred. However, the bound on $B(B\to\rho^0\rho^0)$ is much smaller which indicates that penguin effects are small.  Furthermore, it is dominated by the longitudinal polarization states such that the final state is a CP eigenstate to good approximation.

Even though $\rho^\pm\pi^\mp$ are not CP eigenstates, one can extract $\alpha$ from $B\to\rho\pi$ decays using a Dalitz plot analysis.\cite{Brhopi} The currently best results for $\alpha$ come from this method and from $S_{\rho^+\rho^-}$ leading to the world average\cite{CKM05}
\begin{equation}
  \alpha=\left( 100^{+12}_{-10}\right)^\circ
\end{equation}

\subsection{The angle $\gamma$}
The angle $\gamma$ is the relative phase between $V_{ub}$ and $V_{cb}$ and can therefore be measured when $b\to u$ and $b\to c$ transitions can interfere in one decay. An analysis of the time-dependent CP asymmetry of $B^0\to D^\pm\pi^\mp$ for instance gives $2\beta +\gamma$. Although the final states are not CP eigenstates, the CKM-favored amplitude in $B^0\to D^-\pi^+$ can interfere with the doubly CKM-suppressed amplitude in $B^0\to D^+\pi^-$.\cite{DS} The angle $\gamma$ can be extracted largely independent of new-physics effects from triangle relations between the tree-level decays $B^\pm\to \{D^0, \bar D^0\} K^\pm \to f K^\pm$.\cite{GW,ADS} The classic proposal is by Gronau and Wyler where the $D$ mesons have to be reconstructed as CP eigenstates.\cite{GW} This is experimentally hard because one has to measure small interference terms and is therefore not very sensitive to $\gamma$. Belle overcomes this difficulties using a Dalitz plot analysis of $B^+\to D^0 K^+$ and fits to the interference pattern of $D^0$ and $\bar D^0$.\cite{Bellegam} Not accessible at the current $B$-meson factories are $B_s$ mesons. These are the "eldorado" for $\gamma$ determinations, offering many more possibilities, for example via the decay $B_s\to K^* K^*$. 

\subsection{CKM constraints using theory input}
\begin{figure}
   \begin{center}
	\epsfxsize=7cm\epsffile{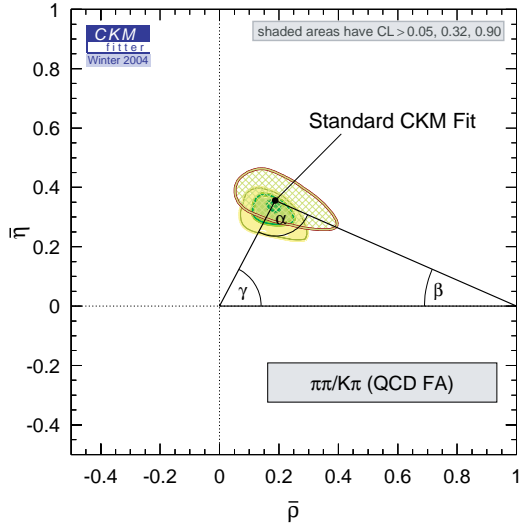}
   \end{center}
	\caption{Fit of QCD-factorization predictions for $B\to\pi\pi/ K\pi$ decays to experimental data from winter 2004. \label{fig:BBNS}}
\end{figure}

Probably the best strategy to extract information on the UT in future will be to use theory input. Fleischer, Gronau, Rosner, and many others suggested to use $SU(2)$ or $SU(3)$ flavour symmetry in addition to some plausible dynamical assumptions for an extraction of $\gamma$.\cite{FGR} Even better control over theoretical uncertainties can be obtained within the framework of QCD factorization where strong phases or the ratio of penguin to tree amplitudes can be calculated.\cite{BBNS} Figure \ref{fig:BBNS} 
shows the potential of using QCD-factorization predictions for non-leptonic charmless two-body $B$ decays to determine $\gamma$. Lacking a recent analysis, the value $\gamma=\left( 62^{+6}_{-9}\right)^\circ$ extracted from data available in winter 2004 is quoted.


\section{Unitarity triangle sides from $B$ decays}
\subsection{The absolute normalization: $|V_{cb}|$}
The CKM matrix element $|V_{cb}|$ can be measured most precisely with inclusive or exclusive semileptonic $b\to c l\nu$ decays. To extract $|V_{cb}|$ from the differential decay spectrum in exclusive $B\to D^{(*)} l\bar\nu$ decays, knowledge of the form factor ${\cal F}_{D^{(*)}}(1)$ at zero recoil is needed. These form factors are genuinely non-perturbative quantities which have to be calculated using e.g. lattice QCD or QCD sum rules. The semileptonic width $\Gamma (B\to X_c l\nu)$ of the inclusive decay, on the other hand, can be calculated using an operator product expansion (OPE) which is a simultaneous expansion in $\alpha_s (m_b)$ and $\Lambda_{\rm QCD}/m_b$. The necessary non-perturbative HQET parameters can be extracted measuring moments of the lepton energy or hadronic invariant mass. A recent global fit to data gives\cite{Vcb}
\begin{equation}
|V_{cb}|=(41.3\pm 0.6\pm 0.1)\cdot 10^{-3}
\end{equation}

\subsection{$|V_{ub}|$ from $B\to X_u l\nu$}
\begin{figure}
   \begin{center}
	\epsfxsize=6cm\epsffile{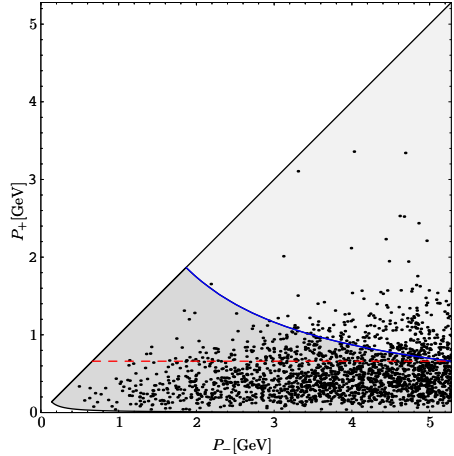}
   \end{center}
	\caption{Hadronic phase space for the light-cone variables $P_-$ and $P_+$. The scatter points indicate the distribution of events. The solid line separates the regions where $s_H<M_D^2$ (dark gray) and $s_H>M_D^2$ (light gray), whereas the dashed line corresponds to $P_+=M_D^2/M_B$.\label{fig:Vub}}
\end{figure}

$|V_{ub}|$ can similarly be extracted from either exclusive or inclusive $b\to u l\nu$ decays. For the inclusive semileptonic $\bar B \to X_u l^- \bar\nu$ decays tight experimental cuts are necessary to discriminate against the large charm background. These cuts restrict the hadronic final state to the shape-function region with large energy $E_X \sim m_B$ but only moderate invariant mass $M_X \sim \sqrt{m_B \Lambda_{\rm QCD}}$. The phase space can be depicted easiest in the hadronic variables $P_\pm = E_H \mp \left| \vec P_H \right|$, the energy of the hadronic final state minus or plus the absolute value of its three momentum.\cite{BLNP} Figure \ref{fig:Vub}
shows the distribution of events in the phase space for the variables $P_-$ and $P_+$. The vast majority of events is located in the shape-function region of small $P_+$ and large $P_-$.

In order to discriminate against the large charm background, one has to apply tight experimental cuts. So far, cuts on the charged-lepton energy $E_{l}>(M_{B}^{2}-M_{D}^{2})/(2M_{B})$, the hadronic invariant-mass squared $s_{H}<M_{D}^{2}$, and the dilepton mass squared $q^{2}>(M_{B}-M_{D})^{2}$ have been employed. Only the $E_{l}$ cut can be applied without neutrino reconstruction. Unfortunately, it has a very low efficiency and is therefore theoretically disfavored. The hadronic-mass cut is in principle the ideal separator between $\bar B\to X_{u} \, l^{-}\bar\nu$  and $\bar B\to X_{c} \,  l^{-}\bar\nu$  events. However, in practice one has to lower the cut due to the experimental resolution on $s_{H}$, thereby reducing the efficiency. Cutting on the hadronic variable $P_+$ provides a new method for a precision measurement of $|V_{ub}|$ which combines good theoretical control with high efficiency and a powerful discrimination against charm background.\cite{BLNP,BLNP2} The fraction $F_P$ of all $\bar B\to X_u l\bar\nu$ events with hadronic light-cone momentum $P_+\leq\Delta_P$ can be calculated systematically using a two-step matching of QCD current correlators onto soft-collinear and heavy-quark effective theory. The prediction for the fraction of events with the optimal cut $P_+\leq M_D^2/M_B$ is\cite{BLNP2}
\begin{equation}
  F_P=(79.6\pm 10.8\pm 6.2 \pm 8.0)\%
\end{equation}
where the errors represent the sensitivity to the shape function, an estimate of ${\cal O}(\alpha_s^2)$ contributions, and power corrections, respectively. The CKM-matrix element $|V_{ub}|$ can be extracted by comparing a measurement of the partial rate $\Gamma_u(P_+\leq\Delta_P)$ with a theoretical prediction for the product of the event fraction $F_P$ and the total inclusive $\bar B\to X_u l\bar\nu$ rate. The resulting theoretical uncertainty on $|V_{ub}|$ is\cite{BLNP2}
\begin{equation}
  \frac{\delta |V_{ub}|}{|V_{ub}|}=(\pm 7\pm 4\pm 5\pm 4)\%
\end{equation}
where the last error comes from the uncertainty in the total rate. At leading power in $\Lambda_{\rm QCD}/m_b$ the shape-function uncertainty could be eliminated by relating the $P_+$ spectrum to the $\bar B\to X_s\gamma$ photon spectrum, both given at tree level directly by the shape function. With a 5\% relative theoretical error on $|V_{ub}|$, corrections suppressed by a power of $\Lambda_{\rm QCD}/m_b$ are considered the second largest source of uncertainty. Subleading shape functions have been investigated first by Bauer, Luke, and Mannel\cite{Bauer:2001mh} and more recently in \cite{Lee:2004ja,Bosch:2004cb,Beneke:2004in} using the two-step matching procedure developed in \cite{BLNP,Bauer:2003pi}. At tree level, the results of this SCET analysis can be expressed in terms of three subleading shape functions defined via the Fourier transforms of forward matrix elements of bi-local light-cone operators in heavy-quark effective theory. 
Numerically, power corrections indeed have the estimated 10\% effect on the value of $F_P(\Delta_P)$ near the charm threshold .

\subsection{$|V_{td}|$ and $|V_{ts}|$}
Finally, we want to discuss the extraction of the CKM matrix elements $|V_{td}|$ and $|V_{ts}|$. The standard way to determine these quantities is via the mass difference $\Delta M_{d/s}$ between the mass eigenstates in the $B_d^0$-$\bar B_d^0$ and $B_s^0$-$\bar B_s^0$ systems, respectively. The quantities
\begin{equation}
  \Delta M_q = \frac{G_F^2}{6\pi^2} \eta_B m_{B_q} B_{B_q} F_{B_q}^2 M_W^2 S_0(x_t) |V_{tq}|^2
\end{equation}
are directly proportional to $|V_{tq}|^2$. In the above formula, $G_F$ is the Fermi constant, $\eta_B=0.55\pm 0.01$ is a QCD factor, $B_{B_q}$ are the so-called bag parameters, $F_{B_q}$ is the $B_q$-meson decay constant, and $S_0(x_t)$ is the Inami-Lim function for the box diagram with a top-quark exchange. Bag parameter and decay constant are non-perturbative input parameters which have to be extracted for instance from lattice QCD data.

$|V_{td}|$ and $|V_{ts}|$ are also accessible in the radiative decays $B\to X_{d/s}\gamma$ which are mediated predominantly via the electromagnetic penguin operator $Q_7$. Whereas the inclusive $B\to X_s\gamma$ branching ratio was measured experimentally to very high accuracy, it is nearly impossible to achieve a corresponding measurement of the inclusive $b\to d\gamma$ branching ratio. The exclusive branching ratios $B\to\rho\gamma$ are experimentally easier accessible. However, they are theoretically more involved than the inclusive modes, because bound state effects have to be taken into account. Yet, a systematic and model-independent analysis of exclusive radiative decays is possible in the heavy-quark limit $m_b\gg\Lambda_{\rm QCD}$. The relevant hadronic matrix elements of local operators in the weak Hamiltonian simplify in this limit because perturbatively calculable hard-scattering kernels can be separated from non-perturbative form factors and universal light-cone distribution amplitudes. This approach to exclusive radiative decays\cite{BFS,BBVgam,AP} is similar in spirit to the treatment of hadronic matrix elements in two-body non-leptonic $B$ decays formulated by Beneke, Buchalla, Neubert, and Sachrajda.\cite{BBNS} The non-perturbative form factors represent the biggest source of theoretical uncertainty in the prediction of the exclusive branching ratios. A very interesting strategy to reduce the form-factor uncertainties is to use ratios like the one of CP-averaged branching ratios
\begin{equation}
  R_0=\frac{B(B^0\to\rho^0\gamma)+B(\bar B^0\to\rho^0\gamma)}{B(B^0\to K^{*0}\gamma)+B(\bar B^0\to \bar K^{*0}\gamma)}
\end{equation}
the isospin breaking ratio
\begin{equation}
  \Delta (\rho\gamma)=\frac{2\Gamma(B^0\to \rho^0\gamma)-\Gamma(B^\pm\to \rho^\pm\gamma)}{2\Gamma(B^0\to \rho^0\gamma)+\Gamma(B^\pm\to \rho^\pm\gamma)}
\end{equation}
or the ratio of $B\to\rho l\nu$ and $B\to\rho\gamma$ decay rates. In these ratios only the ratios of the form factors enter. These form-factor ratios are known in certain limiting cases, like flavour $SU(3)$ or isospin symmetry, or in the large-energy limit.

\begin{figure}
   \begin{center}
	\epsfxsize=6.5cm\epsffile{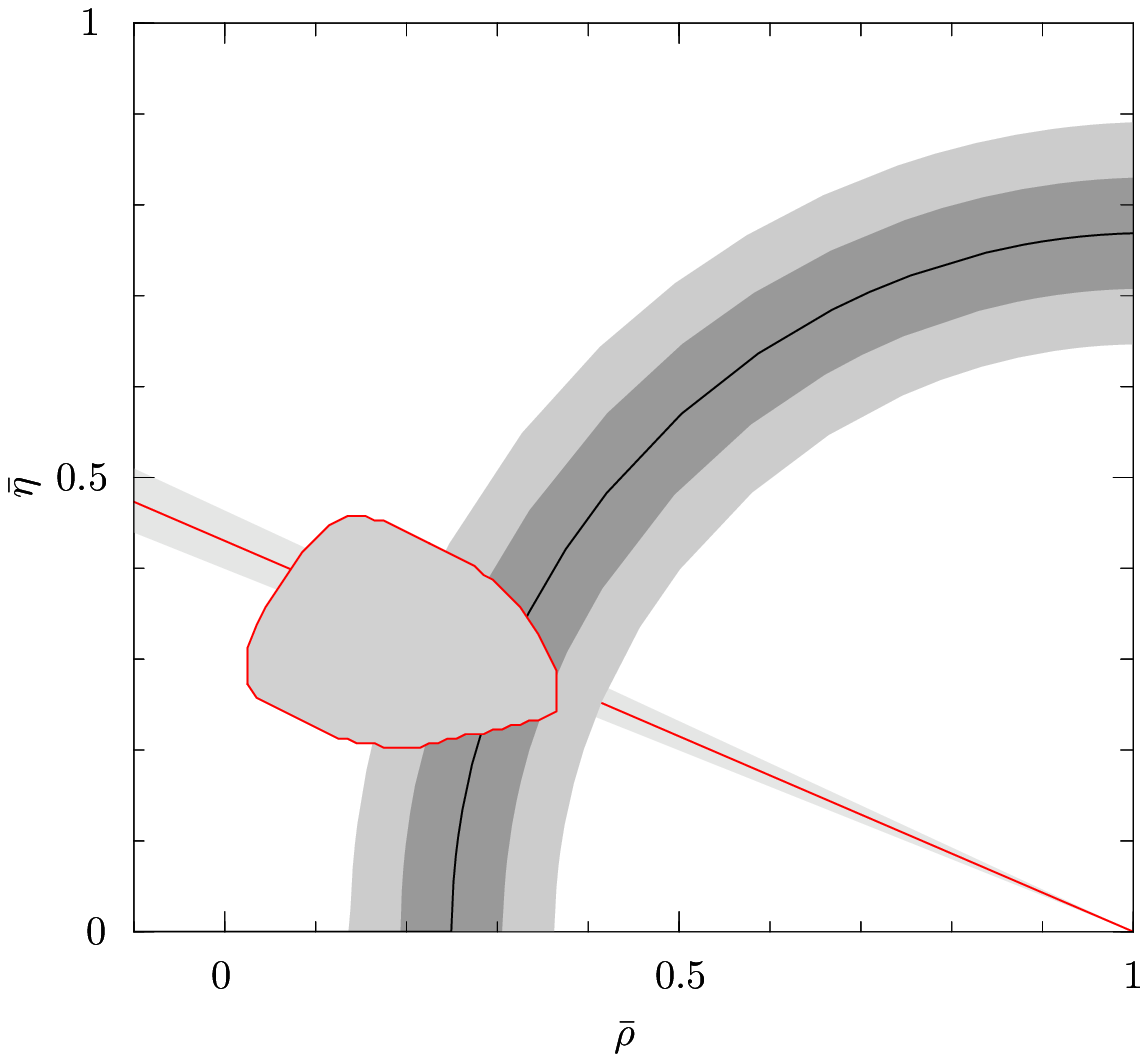}\qquad\qquad
	\epsfxsize=6.5cm\epsffile{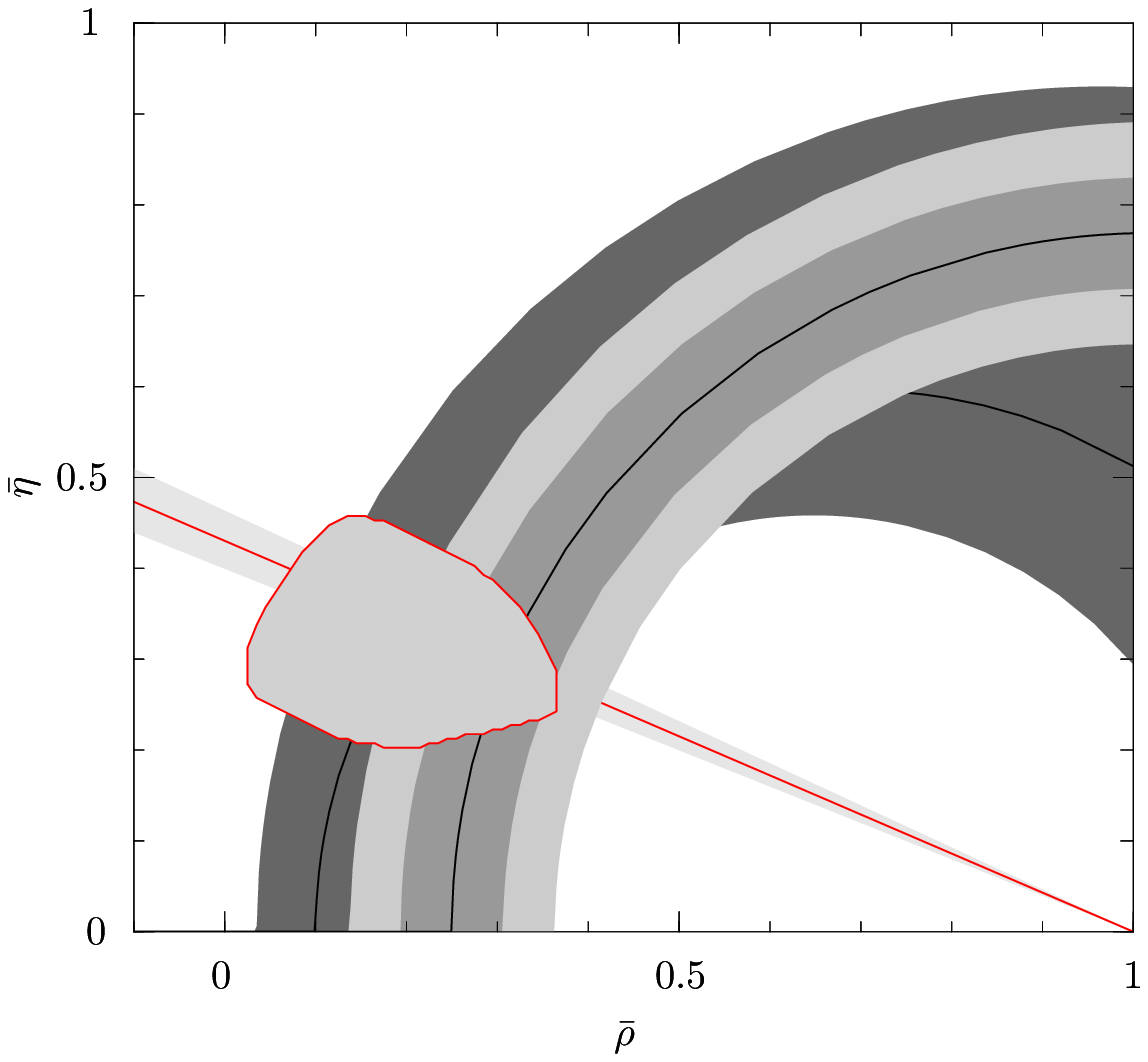}
   \end{center}
   \vspace*{-0.2cm}
	\caption{Left: Constraints implied by $R_0$ in the $(\bar\rho,\bar\eta)$ plane. The width of the dark (light) curved band reflects a 1 (2) $\sigma$ variation of $\xi$ for central $R_0$. The region obtained from a standard fit of the unitarity triangle (irregularly shaped area) and the constraint from $\sin 2\beta$ (light shaded band) are overlaid.
	Right: Same as left plot including the implication of a measurement of $\Delta(\rho\gamma)_{\rm exp}=0$. The width of the band reflects the theoretical uncertainties.\label{fig:R00}}
\end{figure}

The form-factor ratio $\xi=F_{K^*}/F_\rho$ which enters $R_0$ for example differs from unity only because of $SU(3)$-breaking effects. Within the framework of QCD factorization the ratio $R_0$ can be calculated at next-to-leading order in $\alpha_s$ and to leading order in the heavy-quark limit. This ratio measures to very good approximation the side $R_t=\sqrt{(1-\bar\rho)^2+\bar\eta^2}$ of the standard unitarity triangle and the theoretical uncertainty in the relation to $R_t$ comes in essence solely from $\xi$.\cite{ALP,UTBVgam,BFS2} 
From light-cone sum rule and lattice QCD calculations we choose $\xi=1.2\pm 0.1$ as averaged input value for the form-factor ratio $\xi$.\cite{CKM05} The current experimental status of the branching ratio measurements is\cite{BKrhoexp}
\begin{eqnarray}
  B(B^0\to K^{*0}\gamma ) &=& (4.01\pm 0.20)\cdot 10^{-5}\\
  B(B^0\to \rho^0\gamma ) &<& 0.4 \cdot 10^{-6}
\end{eqnarray}
which leads to $R_0<0.01$. This implies
\begin{equation}
  R_t<0.76 \frac{\xi}{1.2},\qquad \left|\frac{V_{td}}{V_{ts}}\right|<0.17 \frac{\xi}{1.2}, \qquad |V_{td}|< 6.7\cdot 10^{-3} \frac{\xi}{1.2}
\end{equation}
The implication of this bound in the $(\bar\rho,\bar\eta)$ plane is shown in the left-hand plot of Figure \ref{fig:R00}. 
We see that the current upper bound for $B(B^0\to\rho^0\gamma)$ already cuts into the standard fit region for the apex of the unitarity triangle. Values of $\bar\rho$ and $\bar\eta$ to the left of the curved band are excluded at 90 \% confidence level. The right-hand plot shows in addition the constraints coming from an assumed measurement of the isospin breaking ratio $\Delta (\rho\gamma)_{\rm exp}=0$, which would correspond to the Standard Model prediction for a CKM angle $\gamma=60^\circ$.

\begin{figure}[t]
\center{\epsfxsize=7cm\epsffile{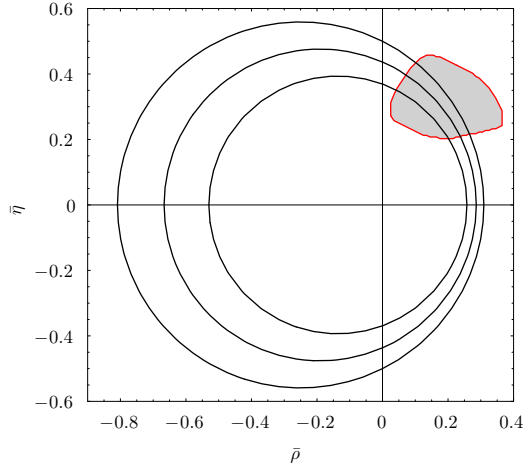}}
\caption{Constraints in the ($\bar\rho,\bar\eta$) plane implied by $|V_{ud}V_{ub}/V_{td}|^2 =0.16\pm 0.04$. For comparison, the standard fit region is indicated by the shaded area.}\label{fig:vubvtd}
\end{figure}

Another way to reduce hadronic uncertainties is to use the ratio of $B\to\rho l\nu$ and $B\to\rho\gamma$ decay rates. The simplification occurs because relations exist between the corresponding form factors in the large-energy limit. Since only $B\to\rho$ transitions are involved, the problems with $SU(3)$ breaking are avoided. The ratio of $B\to\rho l\nu$ events in the part of phase space where the large-energy limit is valid, divided by the $B\to\rho\gamma$ branching ratio then measures the CKM quantity
\begin{equation}
  \left|\frac{V_{ud}V_{ub}}{V_{td}V_{tb}}\right|^2 = \frac{\bar\rho^2 +\bar\eta^2}{(1-\bar\rho)^2+\bar\eta^2}
\end{equation}
With $10^9$ $B\bar B$ pairs available at the end of the Belle and BaBar experiments, a measurement of $|V_{ud}V_{ub}/V_{td}^2|$ with an error of $\pm 25\%$ could be achieved. The corresponding constraint in the $(\bar\rho,\bar\eta)$ plane is shown in Figure \ref{fig:vubvtd}.\cite{UTBVgam}
We observe that the constraint is quite stringent, in particular in the important region corresponding to the standard fit results.

\section{Conclusions}
\begin{figure}
   \begin{center}
	\epsfxsize=7cm\epsffile{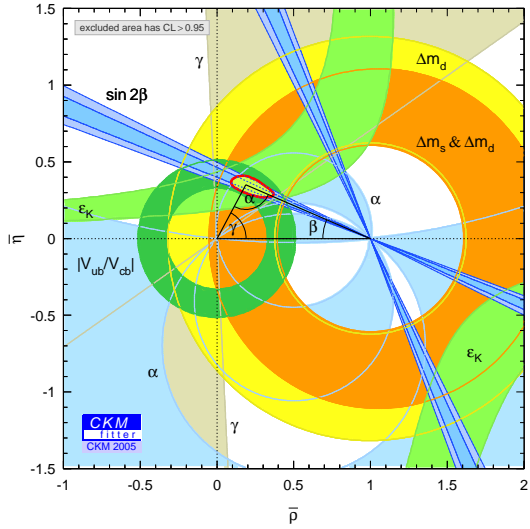}
   \end{center}
	\caption{Global CKM fit in the $(\bar\rho,\bar\eta )$ plane\label{fig:CKMfit}}
\end{figure}

The Standard-Model CKM picture survived its first major test with flying colours: The constraints from the lengths of the sides and those from $\sin 2\beta$ (and $\varepsilon_K$) point to the same region for the apex of the unitarity triangle. This can be seen nicely in Figure \ref{fig:CKMfit} 
which shows the most recent global CKM fit in the $(\bar\rho,\bar\eta )$ plane.\cite{CKMfitter} Now the task is to look for corrections to the standard picture, rather than for alternatives. The only noteworthy discrepancy from theory expectations is the extraction of $\sin 2\beta$ from $b\to s$ penguin modes. If the currently large deviation of $|S_{\psi K_s}-S_{b\to s}|$ from $0$ persists with more data, this would be a rather clean hint towards new physics. The angle $\gamma$ and $|V_{ub}|$ are at the moment the most important theory targets. New extraction techniques for all unitarity triangle parameters are still emerging. During the last years, great progress was made in $B$ physics but no big surprises emerged. The discovery of new physics in the flavour sector seems to be very challenging, both theoretically and experimentally.

\section*{Acknowledgments}
I would like to thank the organizers of the XXXXth Rencontres de Moriond for inviting me to such an enriching conference. Thank you to all the participants for interesting talks, good discussions and great skiing. I acknowledge financial support from the EU Marie Curie Conferences programme. This work was supported in part by the DFG Sonderforschungsbereich/Transregio 9 "Computer-gest\"utzte Theoretische Teilchenphysik". I would like to express my warmest thanks to the Benedictine Archabbey of St. Ottilien for hospitality while these proceedings were completed.\bigskip
\vspace*{0.5cm}
\begin{center}
{\bf --- \quad U.\quad --- \quad I.\quad --- \quad O.\quad --- \quad G.\quad --- \quad D.\quad --- }
\end{center}
\bigskip


\end{document}